\documentclass[aps,prl,superscriptaddress,reprint,floats,citeautoscript,floatfix,nobibnotes,nofootinbib]{revtex4-1}

\usepackage[T1]{fontenc}
\usepackage[intlimits,tbtags]{amsmath}
\usepackage{times}
\usepackage{color}
\usepackage{siunitx}
\usepackage{amssymb}
\usepackage{graphicx}%[draft]
\usepackage{bm}
\usepackage{array}
\usepackage{dcolumn}
\usepackage{bm}
\usepackage{setspace}
\usepackage[T1]{fontenc}
\usepackage[utf8]{inputenc}

\newcolumntype{d}[1]{D{.}{.}{#1} }
\usepackage{xcolor}

\newcommand{\jlua}{International Center for Computational Methods and Software  ${\&}$  State Key Lab of Superhard Materials, College of Physics, Jilin University, Changchun 130012, China.}
\newcommand{\jlub}{International Center of Future Science, Jilin University, Changchun 130012, China.}
\newcommand{\beq}{\begin{equation}}
\newcommand{\eeq}{\end{equation}}
\newcommand{\bea}{\begin{eqnarray}}
\newcommand{\eea}{\end{eqnarray}}
\def\rhor{{\rho({\bf r})}}

\def\br{{\mathbf{r}}}
\def\brp{{\mathbf{r}^{\prime}}}

\begin{document}
\title{Nonlocal Pseudopotential Energy Density Functional for Orbital-Free Density Functional Theory}

\author{Qiang Xu}\affiliation{\jlua}
\author{Cheng Ma}\affiliation{\jlua}
\author{Wenhui Mi}\affiliation{\jlua}
\author{Yanchao Wang}\email{wyc@calypso.cn}\affiliation{\jlua}
\author{Yanming Ma}\affiliation{\jlua}\affiliation{\jlub}

\date{\today}
%\section{Introduction}
%\textbf{Introduction}
\begin{abstract}
Orbital-free density functional theory (OF-DFT) is an electronic structure method with a low computational cost that scales linearly with the number of simulated atoms, making it suitable for large-scale material simulations. It is generally considered that OF-DFT strictly requires the use of local pseudopotentials, rather than orbital-dependent nonlocal pseudopotentials, for the calculation of electron-ion interaction energies, as no orbitals are available. This is unfortunate situation since the nonlocal pseudopotentials are known to give much better transferability and calculation accuracy than local ones. We report here the development of a theoretical scheme that allows the direct use of nonlocal pseudopotentials in OF-DFT. In this scheme, a nonlocal pseudopotential energy density functional is derived by the projection of nonlocal pseudopotential onto the non-interacting density matrix (instead of "orbitals") that can be approximated explicitly as a functional of electron density. Our development defies the belief that nonlocal pseudopotentials are not applicable to OF-DFT, leading to the creation for an alternate theoretical framework of OF-DFT that works superior to the traditional approach.
\end{abstract}

\maketitle

\textbf{Introduction}

$Ab~initio$ calculations using Kohn-Sham (KS) density functional theory (DFT)\cite{HK1964,KS1965} can accurately describe the fundamental properties of various materials. However, its computational cost scales with the cube of the number of electrons in the simulation cell, which poses a major challenge to large-scale simulations. In contrast, orbital-free (OF) DFT is inherently of lower computational cost that scales linearly with the number of atoms in the system, as it relies only on the electron density and the use of KS orbitals is avoided. As a result, OF-DFT is successfully applied to large-scale simulations of systems with up to millions of atoms\cite{cmh2015profess3,cmh2016sbox,sxc2018atlas,sxc2021dftpy}.

The accuracy of OF-DFT simulations depends strongly on the quality of the non-interacting kinetic energy and the electron-ion (or electron-pseudocore) interaction energy employed in the simulations. Many approximate kinetic energy density functionals (KEDFs) have been proposed to evaluate the non-interacting kinetic energy in OF-DFT\cite{eac2002,wya2013,thomas1927,fermi1927,fermi1928,vw1935,hoy1991ksl,perdew1992ksl,thakkar1992ksl,vitos2000ksl,ernzerhof2000ksl,garcia2007ksl,constantin2009ksl,constantin2011ksl,laricchia2011ksl,karasiev2013ksl,constantin2017ksl,lkt2018ksl,constantin2018ksl,lkt2020ksl,cat1985knl,wt1992knl,sm1994knl,perrot1994knl,wgc1998knl,wgc1999knl,garcia2007knl,hc2010knl,constantin2018knl,mgp2018knl,mp2019knl,xwm2019knl,xwm2020knl}. Their use in combination with local pseudopotentials\cite{zbj2004lps,hc2008lps,del2014lps,mwh2016lps,del2017golps} can achieve results that agree reasonably with those derived by KS-DFT, especially for main-group metals, III-V semiconductors\cite{lkt2018ksl,hc2010knl,mgp2018knl,revHC2021}, and even systems with inhomogeneous electron density such as metal clusters and quantum dots\cite{mp2019knl,xwm2020knl}. 

Unfortunately, the local pseudopotentials \cite{wt1992knl,wgc1998knl,wgc1999knl,zbj2004lps,hc2008lps,mwh2016lps} used to evaluate the electron-ion interaction energy suffer from a lack of transferability\cite{mwh2016lps}, as they fail to reproduce the correct scattering behavior of the all-electron potentials\cite{hamann1979ncps,fuchs1999fhipp,martin_2004}. Overcoming the transferability problem requires a reliance on either all-electron potential or nonlocal pseudopotentials (NLPPs), which are widely used in orbital-based approaches. However, it is practically unfeasible to use the all-electron potential, as an accurate all-electron KEDF for OF-DFT calculations is not yet available\cite{lehtomaki2014ofpaw,zavodinsky2019ofaep}.  Furthermore, the use of NLPPs\cite{hamann1979ncps,vanderbilt1990ups} runs against conventional understanding, as no orbitals are available in the traditional framework of OF-DFT\cite{hc2008lps,del2014lps,del2017golps,witt2018ofreview,witt2021ofrandom}. 

A crucial nonlocal energy term with a set of angular-momentum-dependent energies has recently been added to OF-DFT calculations\cite{kyq2013amd1,kyq2014amd2} in an effort to correct errors arising from the use of KEDFs and local pseudopotentials. This approach has successfully reproduced the bulk properties of several standard structures of Ti. However, special care must be taken when applying it to a wide range of practical simulations, as frozen on-site orbitals and empirically directed fitting parameters are part of the model\cite{witt2018ofreview}. There is substantial demand for a general approach to evaluate the electron-ion interaction energy using NLPPs in OF-DFT calculations. In this manuscript, we developed a theoretical scheme that allows the direct use of the NLPPs for the calculation of electron-ion interaction energy in OF-DFT, together with a specially designed theoretical framework of OF-DFT. This development leads to a OF-DFT calculation that gives a better transferability than the existing OF-DFT method based on local pseudopotentials.

\textbf{Results and discussion}

\textbf{Nonlocal pseudopotential energy denstiy functional.} In general, the total energy density functional of OF-DFT can be expressed as:
\bea
E[\rho]&=&T_s[\rho]+E_H[\rho]+E_{XC}[\rho]+E_{II}\left(\{R_{a}\}\right)\nonumber \\ &&+  \int \rhor V_{loc}(\br) d^3\br,\label{eq:1}
\eea
where $\rho$, ${T_s}$, ${E_H}$, ${E_{XC}}$, $E_{II}$, $\{R_{a}\}$, and $V_{loc}$ are the electron density, KEDF, Hartree energy, exchange-correlation energy, ion-ion repulsion energy, the set of atomic positions, and local pseudopotential, respectively.
To include the nonlocal electron-ion interactions, the total energy density functional of OF-DFT is reformulated as:%
\bea
	E[\rho]&=&T_s[\rho]+E_H[\rho]+E_{XC}[\rho]+E_{II}\left(\{R_{a}\}\right)\nonumber \\ &&+\underbrace{\int \rhor V_{loc}(\br) d^3\br+E_{nl}[\rho]}_{E_{EI}[\rho]},\label{eq:2}%
\eea%
where the total electron-ion interaction energy ${E_{EI}[\rho]}$ can be separated in two parts: a local part ${E_{loc}[\rho]=\int{\rho(\br)V_{loc}(\br)}d^3\br}$ and a nonlocal part ${E_{nl}[\rho]}$. All of the terms in Eq.~(\ref{eq:2}) except the nonlocal part of pseudopotential (${E_{nl}}[\rho]$), can be evaluated easily. 

The exact nonlocal pseudopotential energy depends on the KS orbitals or the density matrix:
\begin{eqnarray}
	E_{nl}&\equiv&\sum_{i}f_i\int\int\psi_i^*(\brp)V_{nl}(\brp,\br)\psi_i(\br){d^3\br d^3\brp}\nonumber \\
	&=&\int\int V_{nl}(\brp,\br)\gamma_s(\br,\brp){d^3\br d^3\brp},\label{eq:3}
\end{eqnarray}%
\noindent where $f_i$, $V_{nl}(\brp,\br)=\langle \brp|\hat{V}_{nl}|\br\rangle$, and $\gamma_s(\br,\brp)=\sum_if_i\psi_i(\br)\psi^*_i(\brp)$ represent the occupation number of the $i$-th KS orbital $\psi_i$, the real-space representation of the nonlocal part pseudopotential, and the noninteracting density matrix, respectively. Considering that the density matrices $\gamma_s[\rho](\br,\brp)$  can be used to approximate the KEDFs\cite{eac2002,chakraborty2017kdm1,chakraborty2018kdm1}, a nonlocal pseudopotential energy density functional (NLPPF) relying directly on the density matrix was proposed to evaluate the nonlocal electron-ion interaction energy. The nonlocal electron-ion interaction energy is then rewritten as a functional of electron density%
\begin{eqnarray}
	E_{nl}[\rho]=\int\int{V_{nl}(\brp,\br)\gamma_s[\rho](\br,\brp)}d^3\br d^3\brp.\label{eq:4}
\end{eqnarray}
By taking the Kleinman-Bylander form\cite{kb1982form} of norm-conserving NLPPs, the nonlocal part pseudopotential\cite{tm1991ncpp} can be written as
\begin{eqnarray}
	{V}_{nl}(\brp,\br)=\sum_{a, lm}E_{KB}^{a, lm}\chi_{lm}^a(\brp)\chi_{lm}^{a*}(\br),\label{eq:5}
\end{eqnarray}
where $E_{KB}^{a, lm}=[\int{\phi_{lm}^{a*}(\br)\delta{V}_l^a(\br)\phi_{lm}^a(\br)}d^3\br]^{-1}$ and $\chi_{lm}^a(\br)=\delta{V}_l^a(\br)\phi_{lm}^a(\br)$. The terms $\phi_{lm}^a$ and $\delta{V}_l^a$ are the atomic pseudo-wave-function and the short-range pseudopotential corresponding to the $lm$-th angular momentum of $a$-th atom, respectively. $\gamma_s[\rho]$ denotes the density matrix as a functional of electron distribution $\rho$. Although there is no exact analytic form available for the density matrix functional, a modified Gaussian (MG)\cite{lee1987gdm1} form derived from the second-order Taylor expansions of the density matrix\cite{berkowitz1986dm1} was employed to approximate the density matrix functional:
\beq
\gamma_s^{MG}[\rho](\br,\brp)=\rho(\bar{\br})e^{-\frac{s^2}{2\beta(\bar{\br})} }\left[1+A\left( \frac{s^2}{2\beta(\bar{\br})}\right)^2 \right],\label{eq:6}%
\eeq
where ${s=|\br-\brp|}$ and ${\bar{\br}=(\br+\brp)/2}$. The second term in the square bracket is ${O(s^4)}$ correction\cite{lee1987gdm1}, where ${A}$ is an adjustable parameter. ${\beta(\br)}$ denotes the "local temperature" $\beta(\br)=\frac{3}{2}\frac{\rho(\br)}{t_s(\br)}$\cite{parr1989dft,ghosh1984loct}, where ${t_s(\br)}$ is the exact kinetic energy density defined as ${t_s(\br)}=t^{KS}_s(\br)\equiv\sum_{i=1}^{Occ.}\frac{1}{8}|\nabla\rho_i(\br)|^2/\rho_i(\br)-\frac{1}{8}\nabla^2\rho(\br)$ and ${\rho_i(\br)=|\psi_i(\br)|^2}$  is $i$-th KS orbital's density (see Refs.\cite{berkowitz1986dm1,lee1987gdm1}). To remove the orbital-dependent problem in $t_s(\br)$ and obtain a solely density-dependent form of the density matrix functional, the kinetic energy density is obtained directly from the integrand of KEDFs to replace the exact one: $t_s(\br)\approx t_s[\rho](\br)$. The widely used Wang-Teter (WT) KEDF\cite{wt1992knl} is chosen as an exemplary case, and $t_s[\rho](\br)$ can be expressed as 
\bea
t^{WT}_s[\rho](\br) &=&\frac{3}{10}(3\pi^2)^{2/3}\rho^{5/3}(\br)+\frac{1}{8}\frac{|\nabla\rho(\br)|^2}{\rho(\br)}\nonumber \\ &+&\rho^{5/6}(\br)\int{\omega_{WT}(\br,\brp)}\rho^{5/6}(\brp)d^3\brp,\label{eq:7}
\eea
\noindent where $\omega_{WT}(\br,\brp)$ is the kernel of WT functional. The Supplementary Notes give the details of the kinetic energy densities obtained from WT and Xu-Wang-Ma\cite{xwm2019knl} KEDFs. 

The direct numerical evaluations of $\rho(\bar{\br})$ and $\beta(\bar{\br})$ at the average position $\bar{\br}$ are very complicated. They are therefore approximated using $q$-mean "nonlocal density" $\rho_q(\br,\brp)=\left[\frac{\rho^q(\br)+\rho^q(\brp)}{2}\right]^{1/q}$ and two-point average temperature $\beta(\br,\brp)=[\beta(\br)+\beta(\brp)]/2$ for systems with slowly varying electron densities. The density matrix functional of Eq.~(\ref{eq:6}) can then be reformulated as
\begin{small}
	\begin{equation}
		\tilde{\gamma}_s^{MG}[\rho](\br,\brp)=\rho_q(\br,\brp)e^{-\frac{s^2}{2\beta(\br,\brp)} }\left[1+A\left( \frac{s^2}{2\beta(\br,\brp)}\right)^2 \right].\label{eq:8}%
	\end{equation}
\end{small}
Combining Eqs.~(\ref{eq:4}),~(\ref{eq:5}) and (\ref{eq:8}) gives the NLPPF as
\begin{small}
	\begin{equation}
		E_{nl}[\rho]\approx\sum_{a,lm}E_{KB}^{a,lm}\int_{\Omega_a}\int_{\Omega_a}\chi_{lm}^a(\brp)\chi_{lm}^{a*}(\br)\tilde{\gamma}_s^{MG}[\rho](\br,\brp)d^3\br d^3\brp,\label{eq:9}%
	\end{equation}
\end{small}
\noindent where the integral domain $\Omega_a$ is the $a$-th ionic core region. Owing to the short-range nature of $\{\chi_{lm}^a(\br)\}$, the computational cost of Eq.~(\ref{eq:9}) scales linearly $\mathcal{O}[cN_a]$ with the number of atoms $(N_a)$, where $c$ can be regarded as a constant derived from the double integral within the near-core region. Within the NLPPF scheme, a new theoretical framework of OF-DFT has been built and implemented in ATLAS\cite{mwh2016atlas,sxc2018atlas}. The further computational details are provided in Methods Section. The parameters of pseudopotential and NLPPF are presented in the Supplementary Table 1 and 2, respectively.
\begin{figure*}[!htb]
	\begin{center}
		
		\includegraphics[width=180 mm]{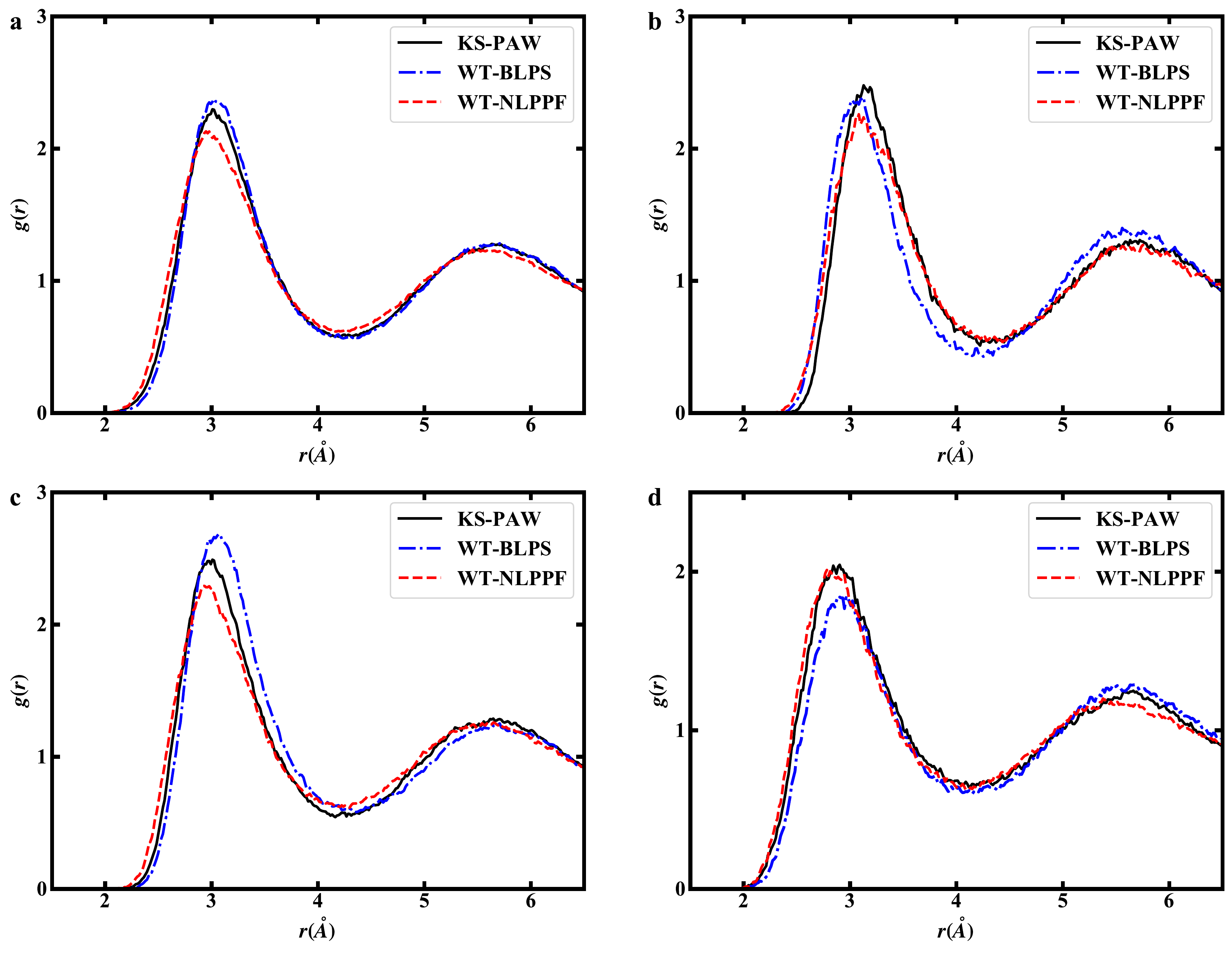}
		\caption{\label{fig:1} \textbf{Pair distribution functions for Li$_{54}$Mg$_{54}$ alloy.} \textbf{a} Total, \textbf{b} Mg-Mg, \textbf{c} Li-Mg, and \textbf{d} Li-Li pair distribution functions. }
	\end{center}
\end{figure*}

\textbf{Computational accuracy of NLPPF scheme.} We first applied this scheme for OF-DFT calculations of Li, Mg, and Cs within hexagonal-close-packed (HCP), face-centered cubic (FCC), body-centered cubic (BCC),simple cubic (SC), and cubic diamond structures. For each structure, 13 energy-volume points were calculated by expanding and compressing the approximate equilibrium volume by up to 20\%, and the bulk properties (the equilibrium cell volume $V_0$, bulk modulus $B_0$, and the relative energy $E_R$ with respect to HCP structure) were determined by fitting the energy-volume curve against Murnaghan's equation of state\cite{murnaghan1924eos}. The comparison of the results obtained by OF-DFT using both local pseudopotentials (BLPS\cite{hc2008lps} and OEPP\cite{mwh2016lps}) and NLPPs against those calculated by KS-DFT using the projector augmented-wave (PAW)\cite{blochl1994paw} method are presented in Supplementary Table~3. For Li and Mg solids, the OF-DFT calculations within both local pseudopotentials and NLPPs give reasonable predictions of $V_0$, $B_0$, and $E_R$, which are comparable with the KS-DFT results. It is noteworthy that our scheme shows an improvement over the local pseudopotentials of OEPP for bulk Cs. The accurate bulk properties of Li/Mg/Cs obtained by the current scheme demonstrate its valid applicability to simple metallic solids.
\begin{figure}[!htb]
	\begin{center}
		\includegraphics[width=88 mm]{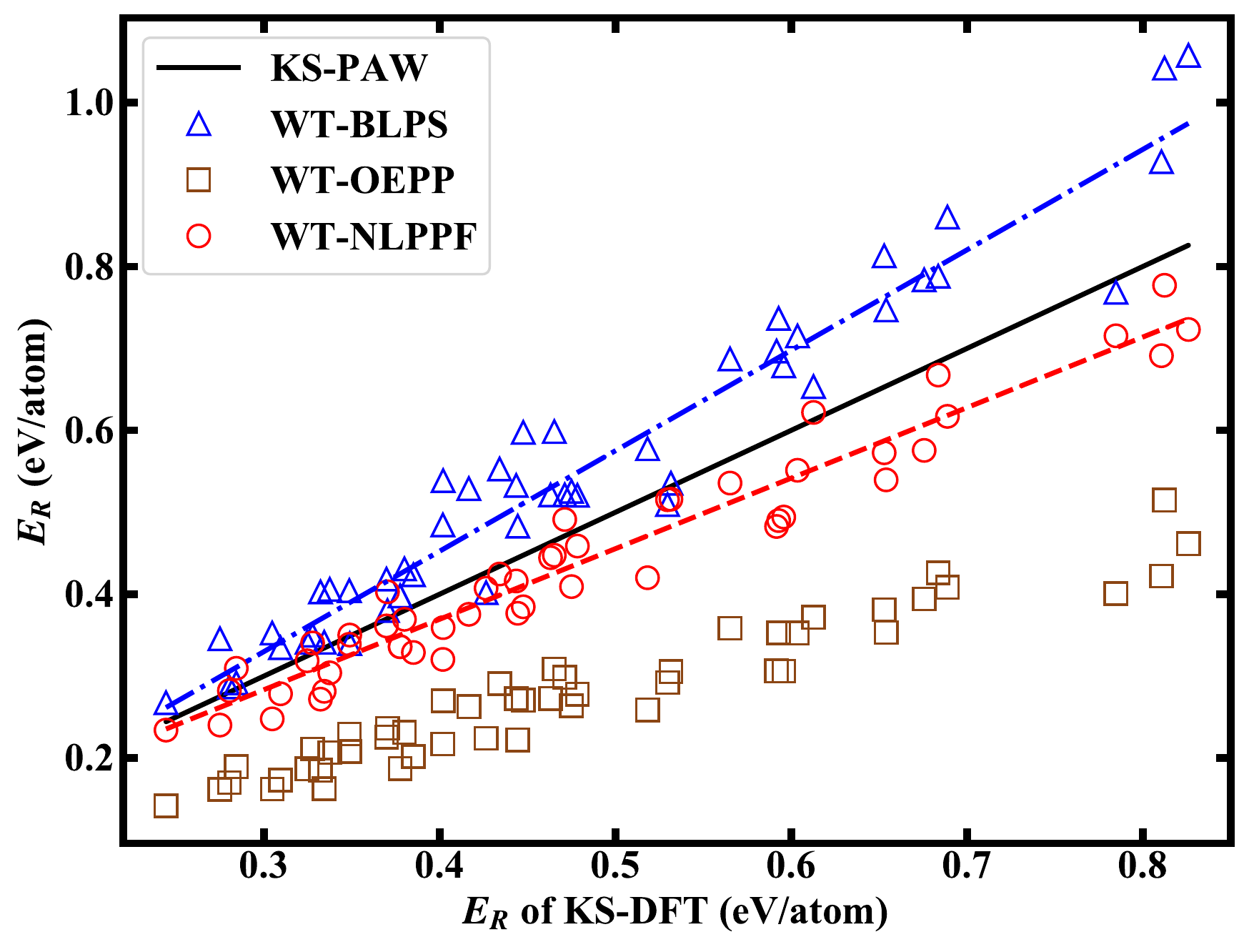}
		
		\caption{\label{fig:2} \textbf{Relative energies for random structures of elemental Li.} The results are calculated by OF-DFT using BLPS, OEPP and NLPPF in comparison with that by KS-DFT using the PAW method. Blue dash-dotted and red dashed lines are the least-square fittings of WT-BLPS and WT-NLPPF results, respectively.}
	\end{center}
\end{figure}
 
 Further assessment of the accuracy of our scheme was demonstrated by molecular dynamics calculations for Li-Mg alloy. The calculations used a canonical ensemble (at 1000K) in a supercell containing 108 atoms (Li$_{54}$Mg$_{54}$). The calculated pair distribution functions $g(r)$ for Li-Mg alloy are shown in Figure~\ref{fig:1}. Overall, the predicted shapes and peak neighbors of pair distribution functions by OF-DFT within NLPPF match the results calculated by KS-DFT. Especially notable are the resulting contributions of the partial distributions (Figures~\ref{fig:1}b-d) calculated by OF-DFT within the NLPPF being almost identical to the KS-DFT calculations, which are superior to those obtained by the local pseudopotentials.

\textbf{Transferability of NLPPF scheme.} To demonstrate the transferability of our scheme, we randomly generated 50 structures of Li systems using CALYPSO\cite{wang2010calypso1,wang2012calypso2}. The total energies of these structures were calculated by OF-DFT and KS-DFT. The comparisons of total energy relative to the HCP structure ($E_R$) are shown in Figure~\ref{fig:2}. The orderings of energy are well captured by OF-DFT within NLPPF. The relative energies of different phases are overall well reproduced and in reasonable agreement with the KS-DFT results. The least-square fitting lines of WT-NLPPF are generally closer to the KS-PAW results than those from the local pseudopotentials. For example, the mean error of $E_R$ for Li systems obtained by OF-DFT within the WT-NLPPF is 45 meV/atom, which is lower than that within BLPS (73 meV/atom), or OEPP (201 meV/atom). Therefore, this framework of OF-DFT with improved transferability is superior to the traditional one.

\begin{table}[!htb]
	\caption{\label{tab:one}$B_0$ (GPa), $E_R$ (eV/atom) and $V_0$ ($\si{\angstrom}^3$/atom) for bulk Li, Mg, and Be by KS-DFT and OF-DFT.}
	\begin{ruledtabular}
		\begin{tabular}{ccccccc}
			&&Method&HCP&FCC&BCC&SC\\ \hline
			Li	&$B_0$&	KS-PAW&	13.9&	13.6&	13.9&	12.1 \\
			& &WT-NLPPF&	13.5&	13.5&13.7&	11.0 \\ 
			&$V_0$&	KS-PAW&	20.280&	20.372&	20.396&	20.580 \\
			& &WT-NLPPF&	19.483&	19.462&	19.352&20.844 \\
			&$E_R$	&KS-PAW&	0.000&	0.000&	0.001&	0.120\\
			& &WT-NLPPF&	0.000&	0.000&0.001&	0.152\\ \hline
			Mg&     $B_0$&  KS-PAW&       35.8&   35.5&   34.8&   22.7\\
			& &WT-NLPPF&  33.0&   31.3&   31.3&   21.2\\
			&$V_0$& KS-PAW&       22.838& 23.071& 22.826& 27.478 \\
			& &WT-NLPPF&  23.194& 23.924& 23.730& 28.274 \\
			&$E_R$& KS-PAW&       0.000&  0.012&  0.029&  0.382 \\ 
			& &WT-NLPPF&  0.000& 0.011& 0.031& 0.372\\ \hline
			Be&	$B_0$&	KS-PAW&	123.3&	119.7&	124.1&	74.5 \\
			&&WT-NLPPF&	91.5&	90.5&	87.2&	63.3 \\
			&$V_0$&	KS-PAW&	7.910&	7.875&	7.822&	10.274\\
			&&WT-NLPPF&	7.690&	7.942&	7.798&	10.160\\ 
			&$E_R$&	KS-PAW&	0.000&	0.080&	0.099&	1.004\\
			&&WT-NLPPF&	0.000&	0.058&	0.082&	0.561
		\end{tabular}
	\end{ruledtabular}
\end{table}

\begin{table}[!htb]
	\caption{\label{tab:two}$B_0$ (GPa), $E_R$ (eV/atom) and $V_0$ ($\si{\angstrom}^3$/atom) for bulk Cd by KS-DFT and OF-DFT.}
	\begin{ruledtabular}
		\begin{tabular}{ccccccc}
			&&Method&HCP&FCC&BCC&SC\\ \hline
			Cd	&$B_0$	&KS-PAW	&40.8	&41.0	&34.5	&29.2	\\
			&&KS-NLPP&67.4	&66.2	&66.1	&39.8	\\
			&&WT-OEPP	&148.2	&138.8	&147.1	&74.8	\\
			&&WT-NLPPF	&67.0	&64.8	&65.2	&39.8	\\ 
			&$V_0$	&KS-PAW	&22.758	&22.979	&23.512	&27.141	\\
			&&KS-NLPP	&15.684	&15.851	&15.674	&18.487	\\
			&&WT-OEPP	&10.379	&10.693	&10.379	&11.964	\\
			&&WT-NLPPF	&15.702	&16.027	&15.763	&18.878	\\ 
			&$E_R$&KS-PAW	&0.000	&0.004	&0.053	&0.121	\\
			&&KS-NLPP	&0.000	&0.019	&0.036	&0.436	\\
			&&WT-OEPP	&0.000	&0.138	&0.072	&0.979	\\
			&&WT-NLPPF	&0.000	&0.036	&0.056	&0.451	\\
		\end{tabular}
	\end{ruledtabular}
\end{table}

\begin{figure*}[t]
	\includegraphics[width=180 mm]{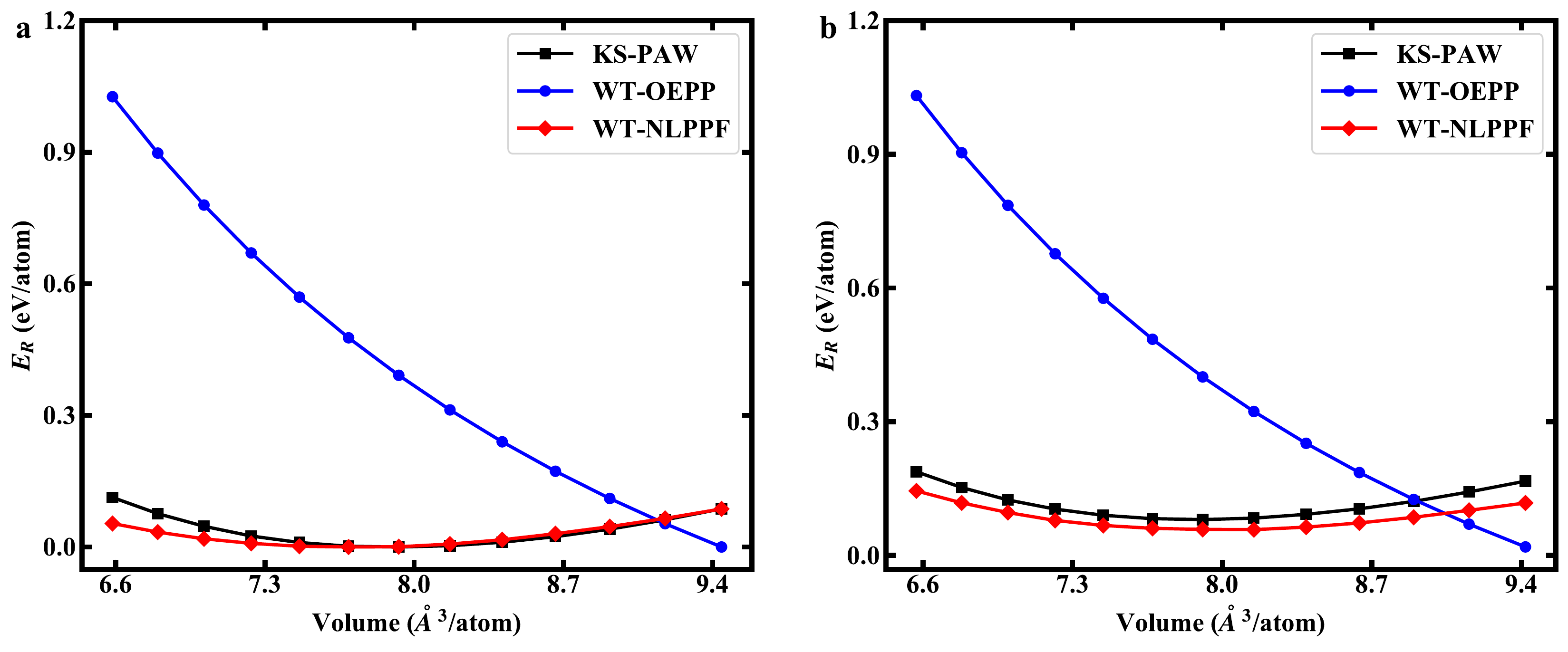}
	\caption{\label{fig:three}\textbf{Relative energy versus volume curves for Be systems.} \textbf{a} The calculated energy-volume curves of Be-HCP. \textbf{b} The calculated energy-volume curves of Be-FCC. The total energy shift of WT-OEPP is -33.030 eV/atom.}
\end{figure*}

\begin{figure*}%[b]
	
	\includegraphics[width=180 mm]{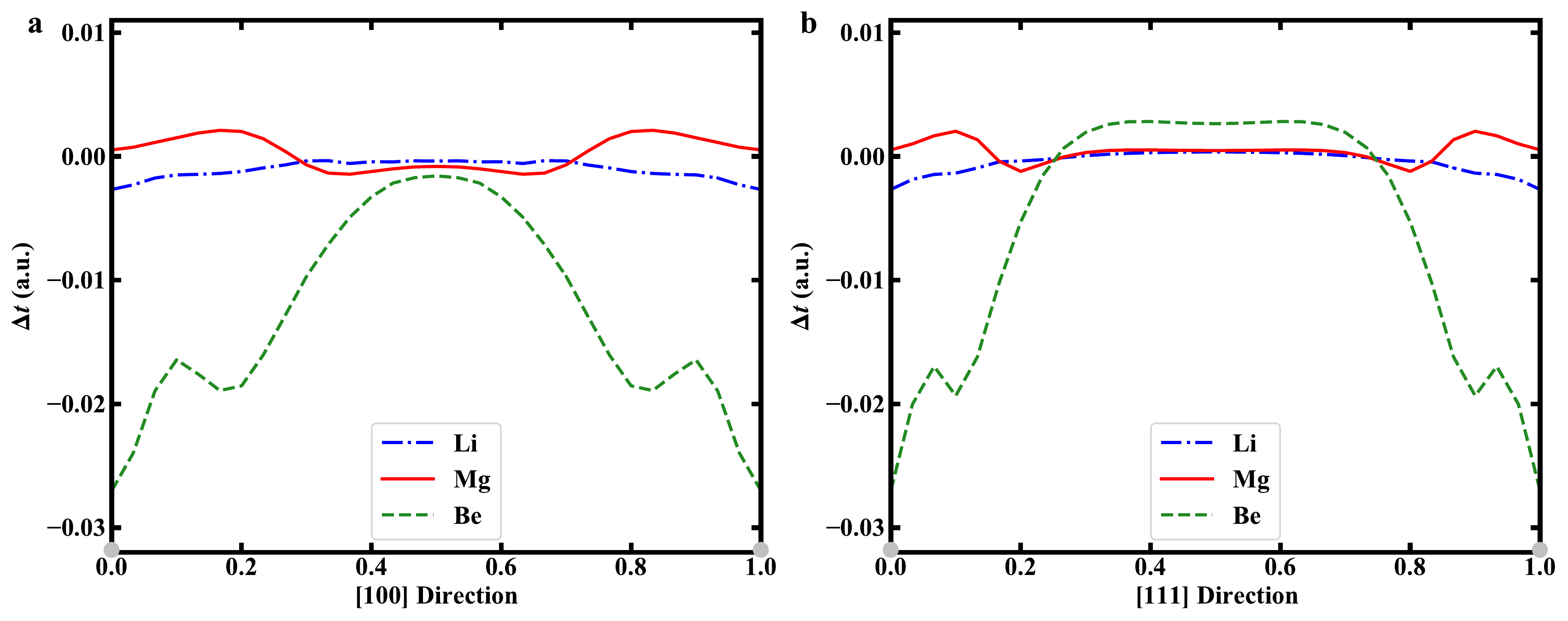}
	\caption{\label{fig:four} \textbf{Kinetic energy differences in SC structure for Li, Mg, and Be systems.}  \textbf{a} The calculated differences along [100] direction. \textbf{b} The calculated differences along [111] direction. Note that $\Delta t(\br)\equiv t^{WT}_s[\rho^{KS}](\br)-t_s^{KS}(\br)$.}
\end{figure*}

Previous studies have shown that OF-DFT with local pseudopotentials can be applied to most $s$- and $p$-block metals. However, OF-DFT simulation using the local pseudopotential OEPP shows unacceptable errors for various crystalline phases of Be  (Figure~\ref{fig:three}): the curves of energy with respect to volume for HCP and FCC structures show the total energy monotonically decreasing with increasing volume. In contrast, the curves with clear minima predicted by OF-DFT within NLPPF agree well with those produced by KS-DFT. These findings indicate the significant superiority of our proposed framework over the conventional one.

Note that the bulk properties (e.g., equilibrium volume, bulk modulus, and relative energy) calculated by OF-DFT within NLPPF reproduce the results of KS-DFT for Li and Mg almost exactly (Table~\ref{tab:one}), considering the maximal deviation of $E_{R}$ is within 32 meV/atom. However, there are some discrepancies for crystalline phases of Be: in particular, the deviation of $E_{R}$ for the SC structure is larger than 400 meV/atom. To explore the causes of these discrepancies, we estimated the errors of the kinetic energy density of the WT-KEDF with respect to the KS kinetic energy density along the [100] and [111] directions in the SC structures of Li, Mg, and Be (Figure~\ref{fig:four}). The kinetic energy density of KS-DFT is clearly reproduced accurately by the WT-KEDF for Li and Mg with slow variations of electron densities. However, it is seriously underestimated for Be, in which the electron distribution rapidly varies in the near-core region. Therefore, we believe that errors in the kinetic energy densities for Be lead to the discrepancy in its bulk properties obtained by the framework of OF-DFT within the NLPPF. The findings are fairly consistent with our expectation that the performance of the NLPPF relies strongly on the accuracy of kinetic energy density, as manifested by Eqs.~(\ref{eq:7}) and (\ref{eq:8}).

Due to existence of significant differences in kinetic energy density between WT and the exact one for Cd systems including localized $d$-channel electrons (see Supplementary Figure 1), the NLPPF using WT cannot be applicable to investigate Cd-systems with $d$-channel electrons. Therefore, the NLPP of Cd was constructed without $d$-channel electrons for the additional calculations. As listed in Table~\ref{tab:two}, the bulk properties predicted by the OF-DFT within NLPPF framework agree fairly well with the predictions by KS-DFT using the same NLPP, whereas WT-OEPP gives serious discrepancies compared with KS-NLPP results in all bulk properties. Although the calculations using NLPP without $d$-channel electrons cannot give the accurate bulk properties for Cd systems (Table 2), OF-DFT within NLPPF works superior to that within OEPP. Overall, it can be expected this NLPPF scheme using accurate KEDF and its kinetic energy density can be applied to the systems including localized electrons, such as the transition metals or covalent systems.

\textbf{Computational efficiency of NLPPF scheme.} To assess the computational efficiency of the current scheme, static simulations of Cs body centered cubic supercells containing 128 to 16,000 atoms were performed by OF-DFT within NLPPF. The total wall time of the single point energy calculations is plotted with respect to the number of atoms in Supplementary Figure~2. The computational cost of this framework clearly scales linearly with the number of atoms in the simulation cell, in sharp contrast to the cubic scaling of KS-DFT. This shows that the OF-DFT within NLPPF is potentially applicable to the simulation of large-scale systems containing millions of atoms.

In summary, we proposed a NLPPF scheme that allows the direct use of NLPPs in OF-DFT calculations. The static and dynamic properties of $s$ and $p$ block metals calculated within this scheme agree well with KS-DFT predictions and show significant improvements on the computational accuracy and transferability over conventional OF-DFT with local pseudopotentials. With this work, we defy the conventional wisdom of orbital-dependent NLPPs being incompatible with OF-DFT, leading to the creation of an alternative framework of OF-DFT, which opens up new avenues for further development of the theory. 

\textbf{Methods}

 \textbf{Pseudopotential generations.} The Troullier-Martins NLPPs\cite{tm1991ncpp} are generated by the FHI98PP\cite{fuchs1999fhipp} code for all considered systems [see Supplementary Table~1] and the $p$-channel of the NLPPs is used as the local pseudopotential of ${V_{loc}(r)}$ in OF-DFT. 
 
 \textbf{Numerical calculations.} The KS-DFT calculations using the PAW\cite{blochl1994paw} and NLPP are performed by VASP\cite{kresse1996vasp1,kresse1996vasp2} and ARES packages\cite{xu2019ares}, respectively. The k-point meshes are generated using the Monkhorst-Pack method\cite{mp1976ksampling} with the k-spacing of 0.10 $\si{\angstrom}^{-1}$. The kinetic energy cutoff is 500 eV for all the simulations using VASP. The OF-DFT calculations are carried out by ATLAS\cite{mwh2016atlas,sxc2018atlas} using WT\cite{wt1992knl} as KEDF, and the corresponding kinetic energy density is used to construct the NLPPF. The generalized gradient approximation with the form of Perdew-Burke-Ernzerhof\cite{pbe1996gga} is employed for both OF-DFT and KS-DFT calculations. The grid spacings of 0.18, 0.18, 0.22, 0.10, 0.15, 0.22, 0.12 and 0.15 $\si{\angstrom}$ are used in ATLAS/ARES for Li, Mg, Cs, Be, Cd, K, Zn and Li-Mg alloy, respectively. The parameters of $A$ and $q$ in NLPPFs are presented in Supplementary Table~2 carefully tuned to yield the bulk properties, which agree with the KS-DFT (NLPPs) predictions. 
 
 \textbf{Molecular dynamics} The molecular dynamic simulations of Li-Mg alloy are performed in the canonical ensemble (at 1000 K) applying the Nos\'e-Hoover thermostat\cite{nose1984md,hoover1985md} simulations up to 10 ps (0.5 fs/step), with the first 10000 steps for equilibrating the system. The data for further analysis were collected from the subsequent 10000 steps.

\textbf{Data Availability}

The authors declare that the main data supporting the findings of this study are contained within the paper and its associated Supplementary Information. All other relevant data are available from the corresponding authors upon reasonable request.
%\bibliography{Refs}
%\bibliographystyle{naturemag}

\textbf{Acknowledgements}
%\begin{acknowledgments}
	Q. X., Y. W. and Y. M. acknowledge funding support from the National Natural Science Foundation of China under Grants No. 12047530, 12034009, 91961204, 11774127, 12174142, 11404128, 11822404, and 11974134; the Program for JLU Science and Technology Innovative Research Team. Part of the calculation was performed in the high-performance computing center of Jilin University. This paper is dedicated to the 70th anniversary of the physics of Jilin University.
%\end{acknowledgments}

\textbf{Author Contributions}

Q. X., Y. W., and Y. M. conceived and designed the project and theoretical framework. Q. X. and C. M. implemented the computational code for this framework in the ATLAS package and performed the computer simulations. Q. X., W. M., Y. W., and Y. M. analyzed the results and wrote the manuscripts. All authors contributed to the discussion and revision of the manuscript.

\textbf{Competing Interests}

The authors declare no competing interests.
\end{document}